\pdfoutput=1

\documentclass[11pt]{article}

\usepackage{amsmath,amsfonts,bm}

\def\eqref#1{equation~\ref{#1}}

\def\1{\bm{1}}

\DeclareMathAlphabet{\mathsfit}{\encodingdefault}{\sfdefault}{m}{sl}
\SetMathAlphabet{\mathsfit}{bold}{\encodingdefault}{\sfdefault}{bx}{n}

\usepackage[final]{acl}
\usepackage{amsmath} 
\usepackage{times}
\usepackage{latexsym}
\usepackage{amssymb}
\usepackage{bbding}
\usepackage{booktabs}

\usepackage[T1]{fontenc}

\usepackage[utf8]{inputenc}

\usepackage{microtype}

\usepackage{inconsolata}
\usepackage{courier}
\usepackage{graphicx}
\usepackage{inconsolata}
\usepackage{colortbl}  
\usepackage{array}   
\usepackage{xcolor}
\usepackage{calc}
\usepackage{float}
\usepackage{pifont}
\usepackage{tabularx}
\usepackage{booktabs}
\usepackage{amsmath}
\usepackage{amssymb}
\usepackage{tcolorbox}
\usepackage{url}
\usepackage{subfigure}
\usepackage{listings}
\usepackage{CJKutf8}
\usepackage{multirow}
\usepackage{arydshln}

\definecolor{darkgreen}{rgb}{0.0, 0.5, 0.0} 
\definecolor{darkred}{rgb}{0.5, 0.0, 0.0}   
\newcommand{\cmark}{\textcolor{darkgreen}{\ding{51}}}%
\newcommand{\xmark}{\textcolor{red}{\ding{55}}}%

\usepackage{CJKutf8}
\usepackage{xcolor, soul}
\definecolor{xcfcolor}{rgb}{0.858, 0.188, 0.478}

\title{MLDebugging: Towards Benchmarking Code Debugging Across Multi-Library Scenarios}

\author{
	Jinyang Huang$^{\spadesuit}$\thanks{\ \ Equal Contribution}
	\quad Xiachong Feng$^{\clubsuit}$\footnotemark[1]
	\quad Qiguang Chen$^{\diamondsuit}$
	\quad Hanjie Zhao$^{\spadesuit}$
	 \\
	\textbf{
		Zihui Cheng$^{\spadesuit}$
		\quad Jiesong Bai$^{\heartsuit}$
		\quad Jingxuan Zhou$^{\spadesuit}$
		\quad Min Li$^{\spadesuit}$
		\quad Libo Qin$^{\spadesuit}$}\thanks{\ \ Corresponding Author}  \\
	$^{\spadesuit}$  School of Computer Science and Engineering, Central South University, China \\
	$^{\clubsuit}$The University of Hong Kong\\
	$^{\diamondsuit}$Research Center for SCIR, Harbin Institute of Technology, Harbin, China\\
	$^{\heartsuit}$ School of Communication and Information Engineering, Shanghai University, China\\
	\texttt{hjy.tsuki@gmail.com}, \texttt{lbqin@csu.edu.cn}\\}

\begin{document}
\maketitle
\begin{abstract}

Code debugging is a crucial task in software engineering, which attracts increasing attention. While remarkable success has been made in the era of large language models (LLMs), current research still focuses on the simple no-library or single-library setting, ignoring the complex multi-library scenario in real-world applications.
To address this limitation, we make the first attempt to introduce MLDebugging (Multi-Library Debugging), a comprehensive benchmark designed to assess debugging challenges within multi-library Python code. Specifically, MLDebugging encompasses 126 distinct Python libraries, covering a wide range of multi-library code issues, categorized into seven distinct types. Furthermore, we conduct a thorough evaluation of MLDebugging using both mainstream open-source and closed-source LLMs and highlight that current LLMs still struggle to correctly perform code debugging across multi-library scenarios. We hope this work can uncover the potential of LLMs in multi-library debugging scenario and offer insights for future research.

\end{abstract}
\section{Introduction}
Code debugging emerges a significant urge for code review, requiring bug location first and then fixing the bug for correct functionality, which has garnered increasing attention for software engineering~\citep{just2014defects4j,lin2017quixbugs}.
In light of the need to enhance the efficiency of code debugging and repair, a series of work consider adapting Automatic Code Debugging (ACD) techniques to serve as a fast and promising solution to the persistent issue of software defects~\citep{austin2021program,chen2021evaluating,li2024codetree,shi2024code}.
\begin{figure}[t]
	\includegraphics[width=0.98\columnwidth]{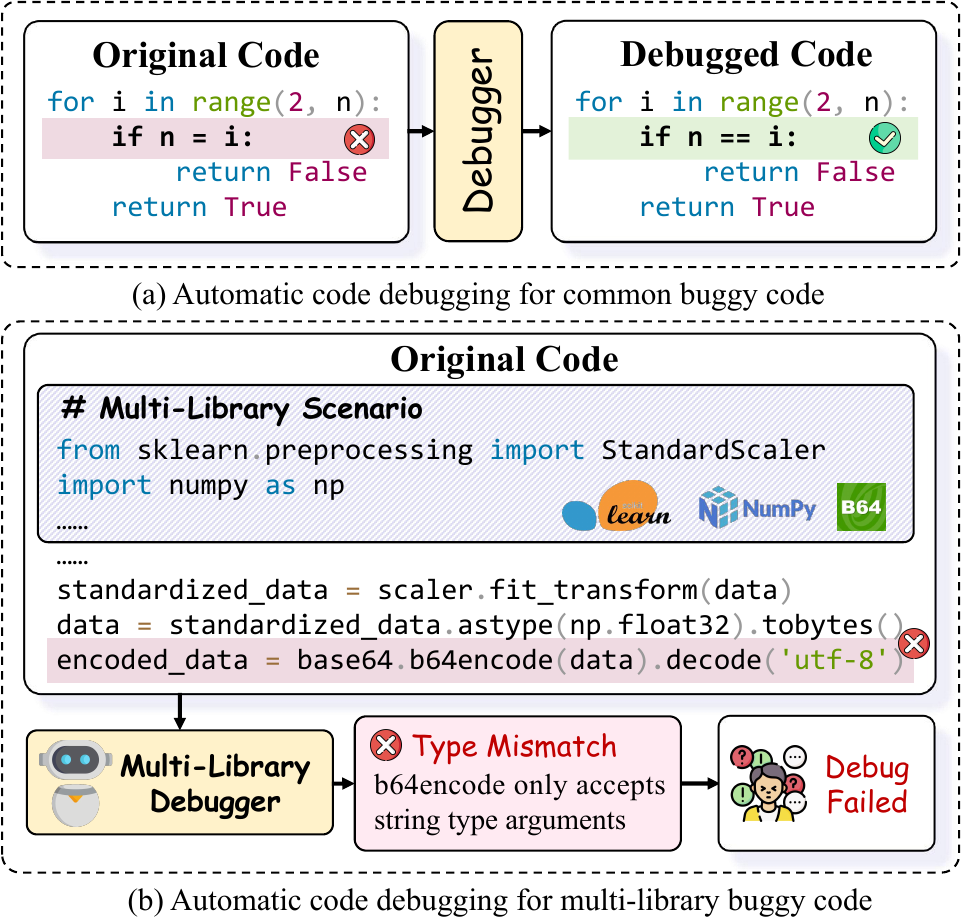}
	\caption{A representative static bug example (a) alongside our newly introduced multi-library bug instance (b) is shown. In (a), the error arises from using the assignment operator '=' instead of the equality comparison operator '==', while (b) involves an issue of variable adaptation between two library functions.}
	\label{fig:intro}
\end{figure}
\begin{table*}[t]
	\footnotesize
	\centering
	\begin{tabular}{l  lccc}
		\toprule
		Benchmark & Language Type & Libraries & Reference Code & Task Scenario \\
		\midrule
		xCodeEval   & 11 types         & [0, 1] & \cmark & Algorithm Competition     \\
		HumanEval    & Python           & [0, 1] & \cmark & Simple Requirements       \\
		MdEval       & 18 types         & [0, 1] & \cmark & Simple Requirements       \\
		QuickBugs    & Python, Java     & [0, 1] & \xmark & Algorithm Competition     \\
		\midrule
		Ours  & Python           & [2, 6] & \cmark & 7 Real-World Scenarios    \\
		\bottomrule
	\end{tabular}
	\caption{Comparison of representative code debugging and generation benchmarks.}
	\label{tab:benchmark-comparison}
\end{table*}


With the advancement of large language models (LLMs) \citep{zhao2025surveylargelanguagemodels, qin2024largelanguagemodelsmeet}, a considerable body of research has been dedicated to effective code debugging. Specifically, \citet{berabi2021tfix} was the first to reframe the code debugging task as a Text-to-Text problem. \citet{tian2024debugbench} introduced the first debugging dataset specifically designed for LLMs, which leveraged code snippets from the \citet{leetcode2025} platform. \citet{khan2024xcodeeval} proposed multiple debugging sub-tasks, thereby expanding the code debugging task to encompass multi-language and multi-task scenarios.
Furthermore, \citet{liu2024mdeval} further extended this benchmark to include multilingual debugging contexts, covering 18 programming languages. This extension facilitates a more thorough assessment of LLM debugging capabilities across various programming languages.

Despite its success, as shown in Figure~\ref{fig:intro} (a), the current research are still limited to the simple no-library or single-library setting, which fails to satisfy the requirements of some complex multi-library scenario in  real-world applications.
 
Actually, 

in real-world software development, the use of multiple libraries is a common practice, as evidenced by research~\citep{feng2024complexcodeeval}, which emphasizes the importance of multi-library scenarios for code debugging. Unlike the previous no-library or single-library scenarios, 

as illustrated in Figure~\ref{fig:intro} (b), multi-library debugging natrually introduce two distinct challenges:  
(1) \textit{Understanding multiple libraries for bug location} and (2) \textit{Utilizing multiple libraries for bug fixing}, which cannot be addressed by previous approaches.

Driven by this motivation, we propose MLDebugging (Multi-library Debugging) in this study, a benchmark designed to evaluate debugging across 126 libraries, comprising 1,175 samples. As shown in Table \ref{tab:benchmark-comparison}, the task involves providing a code snippet that integrates multiple libraries, along with descriptions of the required functionality, test cases, and reference code. We employ GPT-4o \citep{GPT-4o} to produce buggy code samples derived from the multi-library code generation benchmark \citep{zhuo2024bigcodebench}, which are then debugged through the application of multiple LLMs. Next, we design a bug category balancing process, enabling the generation of more stable and balanced bugs. Finally, we implement rigorous quality control to measure and validate the quality and authenticity of our dataset by comparing it with the distribution of real-world multi-library bugs.

To assess the limitations of current LLMs, we carry out a comprehensive evaluation of both open- and closed-source LLMs using MLDebugging. Our experiments reveal the following insights:
\textit{(1) Current LLMs excel at debugging method class errors but struggle with conceptual mistakes and missing imports.
(2) The structured nature of the MLDebugging, widespread use of libraries, and access to complete runtime information, such as test cases and feedback, enhance LLM performance.
(3) In MLDebugging, models like DeepSeek-r1 \citep{guo2025deepseek}, which are built on distillation techniques, fall short of enhancing task performance.}

Our contributions are summarized as follows: 
\begin{itemize}
	\item [(1)] We introduce a complex scenario of multi-library code debugging, addressing challenges encountered in real-world development tasks.
	\item [(2)] We construct a multi-library code debugging benchmark with a total of 1,175 samples, covering 126 commonly used software libraries and categorized into 7 distinct bug types relevant to multi-library environments.
	\item [(3)] We conduct a comprehensive analysis of the dataset's performance across multiple models and provide detailed insights, with further exploration following the experiment.
\end{itemize}

To promote further investigation, the complete dataset can be accessed via \url{https://github.com/hjyTsuki/MLDebugging}.

\section{Task Formulation}
Consider a complete library set \( \mathcal{L} \), an error code  \( C_{\mathcal{R},l} \) that implements a particular requirement $\mathcal{R}$, and utilizes the subset of libraries \( l \subseteq \mathcal{L} \). Given an ideal test case set \( \mathcal{T} \), the error code \( C_{\mathcal{R},l} \) is defined to satisfy the following condition:
\begin{equation}
	\exists t\in \mathcal{T}, \texttt{exec}(C_{\mathcal{R},l}|t) = \texttt{error},
\end{equation}
where $\texttt{exec}(x|y)$ denotes the execution of code $x$ with input-output assertion $y$, returning \texttt{error} if execution encounters a fault, or \texttt{pass} if the execution is successful without errors.

In the context of multi-library code debugging, the task involves generating the correct code \( \hat{C}_{\mathcal{R},l} \) from the erroneous code \( C_{\mathcal{R},l} \). This process can be formally expressed as:
\begin{equation}
	\hat{C}_{\mathcal{R},l} = \mathcal{D}(C|C_{\mathcal{R},l}, \mathcal{R}, \mathcal{L}),
\end{equation}
where \( \mathcal{D} \) represents the debugger, which utilizes the library set \( \mathcal{L} \) to correct the original code \( C_{\mathcal{R},l} \). The resulting corrected code $\hat{C}_{\mathcal{R},l}$ will pass all test cases, satisfying the following condition:
\begin{equation}
	\forall t \in \mathcal{T}, \texttt{exec}(C_{\mathcal{R},l}|t) = \texttt{pass}.
\end{equation}



\section{Data Collection}

\begin{figure*}[t]
	\includegraphics[scale=0.5]{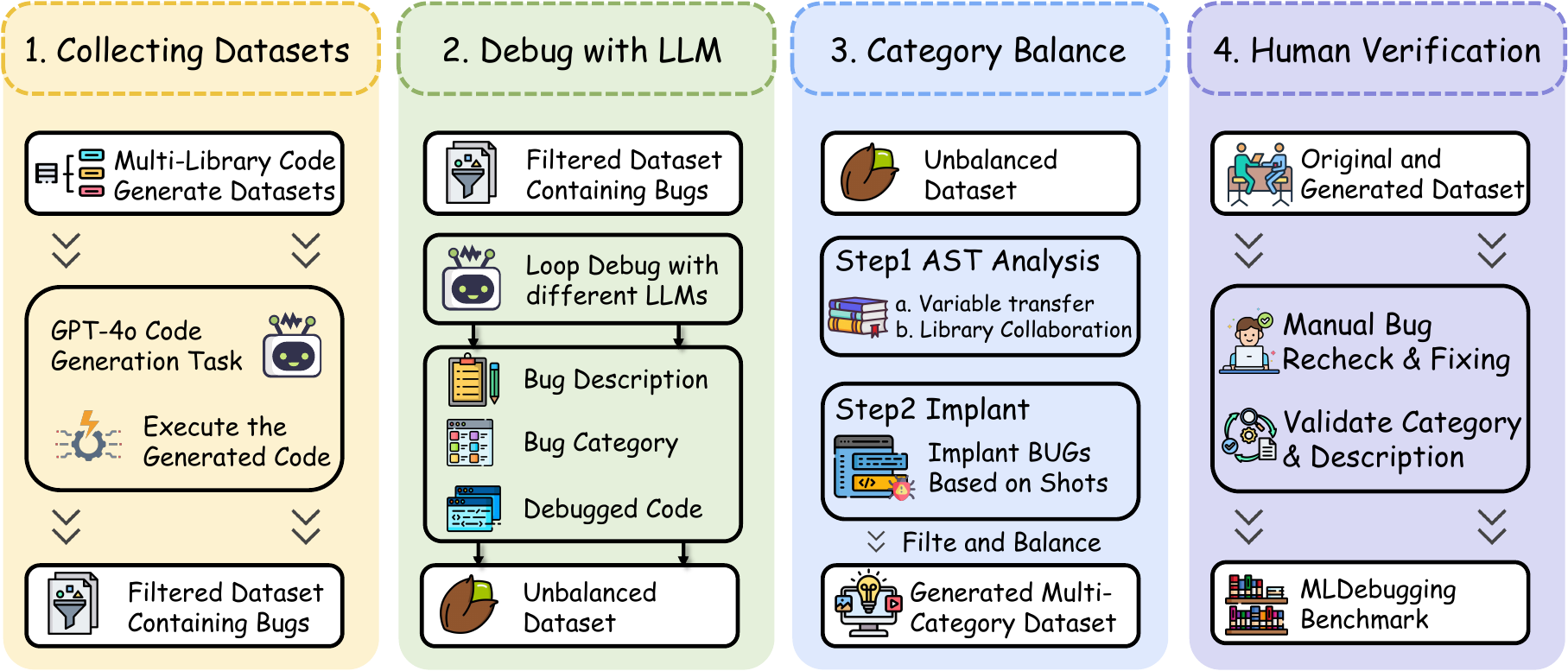}
	\caption {A pipeline diagram illustrating the dataset construction process.}
	\label{fig:pipeline}

\end{figure*}
\begin{table}
	\small
		\centering
		\begin{tabular}{l r}
			\toprule
			\textbf{Type} & \textbf{Count} \\
			\midrule
			Type Mismatch (TM) & 97 \\ 
			Data Transfer Issues (DTI) & 127 \\ 
			Function Parameter Errors (FPE) & 88 \\ 
			Parameter Configuration Errors (PCE) & 60 \\ 
			Function Misuse (FM) & 101 \\ 
			Requirement Misunderstanding (RM) & 143 \\ 
			Import Errors (IE) & 23 \\ 
			\bottomrule
		\end{tabular}
		\caption{The count of samples per category in the \textbf{unbalanced dataset}. In the subsequent sections of the article, we use the first letter shortened forms to replace the full names of the categories (TM, DTI, PFE, PCE, FM, RM, IE).}
			\label{tab:count}
	\end{table}
	


\subsection{Source Code Collection}
\label{sec:prepare}
To obtain realistic erroneous code, as shown in Figure~\ref{fig:pipeline}, we collect practical source code errors as follows:
(1) First, to collect queries that involve multiple libraries, we adapt BigCodeBench~\cite{zhuo2024bigcodebench}, a dataset consisting of code snippets that address real-world programming tasks using two or more Python libraries selected from a pool of 179 widely-used libraries.
(2) Next, we leverage GPT-4o \citep{GPT-4o} following the methodology in \citet{zhuo2024bigcodebench} to generate 1,038 code snippets involving multiple libraries, ensuring a broad range of real and diverse code errors.
(3) Finally, we test all these generated code snippets using the provided test cases, which successfully identifies 609 buggy code snippets.

To further enable a thorough analysis of the dataset, we examine a set of common multi-library bugs preliminary. As shown in Table~\ref{tab:count}, we classify these bugs into 7 categories, each based on one of three perspectives: variable transfer between libraries, library function parameters, and Functionality comprehension.
Based on this analysis, we introduce a clearer and more analyzable classification framework for evaluating debugging in practice, enabling more precise assessments of debugging performance across different bug types.

\subsection{Annotating \& Debugging With LLM}
\label{sec:annotation-llm}

As shown in Figure~\ref{fig:pipeline}, based on the previous classification, we manually provide detailed descriptions and examples for each bug category.
Using this bug category information, we instruct the LLMs to classify each bug and generate a detailed bug description for each code snippet. This process is designed to assist and accelerate human annotation and next model debugging.

To enhance diversity and improve debugging performance, we employ three LLMs: GPT-4o~\citep{GPT-4o}, DeepSeek-V3~\citep{liu2024deepseek} and Claude-3-5-sonnet~\citep{anthropic2024claude}, so that we systematically collected the corrected code results for subsequent comparative analysis.
However, it is worth noting that LLMs cannot always generate and repair code correctly on the first attempt. Inspired by the idea of test-time scaling~\cite{wu2024inference,chen2025ecm}, for any unsuccessful debugging attempts, we conduct up to 5 additional trials to obtain a correct repaired code.  After that, we obtain an unbalanced dataset, each comprising the bug category, corresponding correct and erroneous code pairs, and relevant test cases.

\subsection{Bug Category Balance}

As shown in Table~\ref{tab:count}, an imbalance in bug category distribution leads to evaluation bias, particularly for less frequent errors. To address this, as illustrated in Figure~\ref{fig:pipeline}, we employ a balancing strategy.

\paragraph{Multi-Library Information Preparation}
Analyzing source code in isolation often fails to capture the abstract semantics of code that incorporates multiple libraries, limiting a deeper understanding of its functionality and hindering the generation of debugged code.
To address this, We utilize the Abstract Syntax Tree (AST) to capture the structure of code.
Specifically, we prompt the LLM with the AST, ensuring it reflects variable transfers between libraries, the role of each library at each step, and how they collaborate to accomplish the task.

\paragraph{Category Balance}
Based on the Information Preparation for Multi-Library code, we select a specific bug type from Table \ref{tab:count} and randomly extract corresponding bug instances from the unbalanced dataset.
Specifically, we equally sample each unbalanced category to generate more code with bugs and automatically generate debugged code based on strategies in Sec.~\ref{sec:prepare} and Sec.~\ref{sec:annotation-llm}. 
Next, we manually filter the generated samples, and finally keep sample size of all categories is left are close. This methodology allowed for the successful injection of 566 bugs, standardizing the number of instances per category to approximately 200.

\subsection{Quality Control}

To ensure the integrity of the MLDebugging dataset, we conduct thorough manual quality checks across all data instances.

\paragraph{Manual Bug Recheck \& Fixing}
Due to the model's inability to resolve all bugs, a manual review and correction process is employed for the unresolved code.
Specifically, 4 experienced programmers, each with over 4 years of coding experience, are assigned to the task of bug fixing. Prior to beginning their work, these programmers undergo training on 50 sample cases to ensure consistency in labeling and to standardize the review process.
To ensure the reliability of the bug-fixing process, overlapping cross-checks are organized, allowing for multiple reviews of the same cases. Any discrepancies identified during these reviews are resolved through collaborative discussions, ensuring accuracy and consistency in the final corrections.

\paragraph{Category \& Description Recheck}
Finally, we manually assess the correctness of bug categories and bug descriptions to ensure that MLDebugging provides a reliable assessment of the model’s effectiveness in debugging different categories of bugs. Specifically, this process involves annotators comparing the fixed code with the buggy version to validate the correctness of the generated bug category and bug descriptions, as discussed in Sec.~\ref{sec:annotation-llm}. 

\begin{table}
		\centering
		\begin{tabular}{l r}
			\toprule
			\textbf{Criteria} & \textbf{Correction Count} \\
			\midrule
			BUG Description & 119 \\ 
			BUG Type & 340 \\ 
			  Manual Debugging & 185 \\ 
			\bottomrule
		\end{tabular}
		\caption{The number of erroneous samples identified through manual inspection at the quality control threshold.}
		\label{tab:qulity}
	\end{table}

As a result of these efforts, as detailed in Table \ref{tab:qulity}, we corrected 119 bug descriptions, 340 classifications of multi-library bugs, and manually fixed 185 buggy samples. Additionally, we removed 356 unreasonable samples from the generated dataset. 

\section{Dataset Analysis}
In this study, we design distribution-based experiments to evaluate the alignment between error distributions in our dataset and real-world debugging scenarios. First, we extract question-answer pairs focused on issues related to multiple libraries from \citet{stackoverflow}. We then apply textual embeddings to the error descriptions in both the MLDebugging and DebugBench datasets.

To quantify the distributional similarity, we use two key measures:
(1) \textbf{Centroid-Based Comparison} We calculate the centroids of bug description embedding vectors for MLDebugging, DebugBench, and StackOverflow separately, then evaluate the cosine similarity and Euclidean distance between them.
(2) \textbf{Real-World Proximity} For each sample, we measure its distance to the nearest real-world sample from StackOverflow. Consider the text embedding vector $\mathbf{a}$ from the benchmark (MLDebugging, DebugBench) and the real bug description embedding vector $\mathbf{b}$ from StackOverflow. We retain the $\mathbf{b_j}$ points closest to $\mathbf{a_i}$ and compute the sum of their Euclidean distances to obtain the Distribution Distance (DD) as:
\begin{equation}
\text{DD} = \sum_{i=1}^{m} \min_{j} \|\mathbf{a}_i - \mathbf{b}_j\|_2.
\end{equation}

\begin{figure}[t]
	\includegraphics[width=\columnwidth]{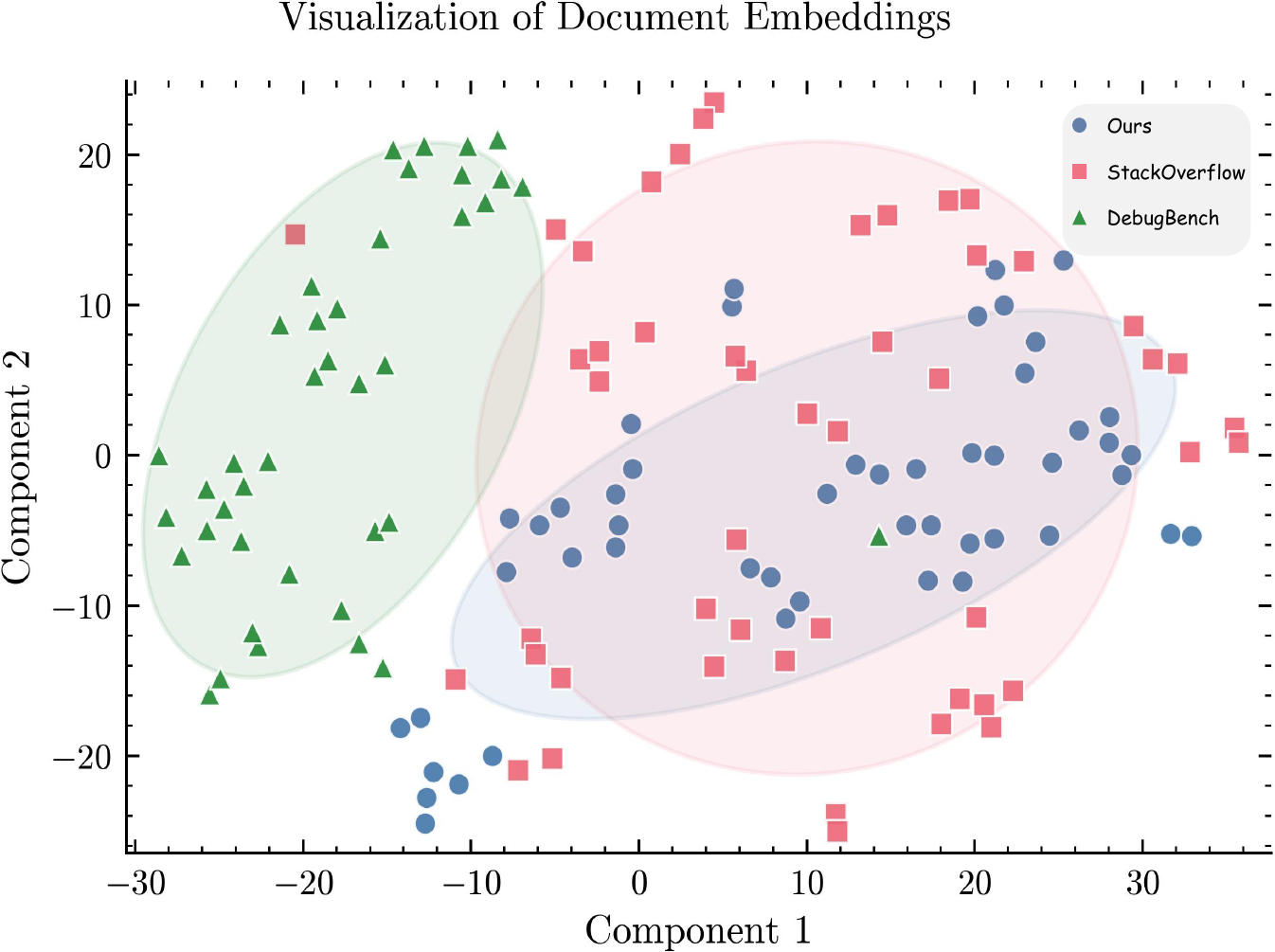}
	\caption{The t-SNE visualization of document embeddings for three datasets in a two-dimensional space. }
	\label{fig:distribution}
\end{figure}

The results in Table~\ref{tab:compare} show that the cosine similarity between our dataset and StackOverflow exceeds that between DebugBench and StackOverflow. Additionally, the overall distances for our dataset are smaller than those for DebugBench, indicating that, at the script level, MLDebugging more accurately reflects the real-world bug distribution on StackOverflow. Moreover, as shown in Figure~\ref{fig:distribution}, the dimensionality reduction visualizations further reveal that our dataset forms more compact clusters, more closely aligning with real-world samples and emphasizing its practical relevance for multi-library code debugging.
\begin{table}[t]
	\centering
	\begin{tabular}{l l l r}
		\toprule
		\textbf{Comparison} & \textbf{C Sim}($\uparrow$) & \textbf{E Dist}($\downarrow$) & \textbf{DD}($\downarrow$) \\
		\midrule
		ML and ST & 0.731 & 0.376 & 46.68 \\
            DB and ST & 0.660 & 0.432  & 56.46  \\
		\bottomrule
	\end{tabular}
	\caption{Comparison of the distances among MLDebugging (ML), DebugBench (DB), and StackOverFlow (ST), where \textbf{C Sim} represents the cosine similarity, and \textbf{E Dist} represents the Euclidean distance. 
	}
    \label{tab:compare}
\end{table}

\section{Experiment}

\begin{table*}[h]
    \centering
    \scalebox{0.7}{
    \begin{tabular}{lcccccccccc}
    
        \toprule  
            \multirow{2}{*}{\textbf{Category}}
            & \multicolumn{6}{c}{\textbf{7B+}} 
            & \multicolumn{4}{c}{\textbf{14B+}} \\
            \cmidrule(r){2-7} \cmidrule{8-11} 
           & \textit{Qwen2.5} & \textit{Qwen2.5-coder} & \textit{Llama3.1} & \textit{Mistral} & \textit{DS Qwen} & \textit{DS Llama} & \textit{Qwen2.5} & \textit{Qwen2.5-coder} & \textit{DS Qwen} & \textit{DS-coder-Lite} \\
        \midrule
TM & 47.6 & 40.0 & 39.7 & 28.8 & 18.8 & 33.5 & 50.0 & 54.1 & 42.4 & 30.0 \\
DTI & 36.1 & 33.8 & 30.5 & 22.7 & 15.7 & 20.5 & 40.7 & 43.1 & 30.6 & 25.5 \\
PFE & 48.4 & 48.8 & 43.2 & 29.1 & 23.5 & 26.2 & 56.8 & 62.0 & 49.8 & 34.7 \\
PCE & 57.6 & 58.0 & 49.8 & 42.4 & 33.1 & 43.0 & 66.1 & 63.8 & 57.6 & 40.1 \\
FM & 38.2 & 40.4 & 38.8 & 26.8 & 20.2 & 23.0 & 44.1 & 53.6 & 35.5 & 31.7 \\
RM & 12.6 & 7.0 & 5.6 & 8.4 & 2.8 & 7.5 & 15.4 & 16.1 & 11.2 & 4.9 \\
IE & 26.1 & 8.7 & 13.0 & 4.3 & 3.3 & 19.0 & 30.4 & 30.0 & 17.4 & 17.4 \\
\hdashline
\textbf{AVG.} & \textbf{42.7} & 40.6 & 36.7 & 28.0 & 20.6 & 27.6 & 48.6 & \textbf{51.4} & 40.2 & 29.9 \\
    \toprule  
    
     \multirow{2}{*}{\textbf{Category}} & \multicolumn{4}{c}{\textbf{32B+}}
     & \multicolumn{2}{c}{\textbf{72B}}
    & \multicolumn{4}{c}{\textbf{MOE}}\\
    \cmidrule(r){2-5} 
    \cmidrule(r){6-7} 
    \cmidrule{8-11} 
    \textbf{} & \textit{Qwen2.5} & \textit{Qwen2.5-Coder} & \textit{DS Qwen} & \textit{QwQ} & \textit{Qwen2.5} & \textit{Llama3.1} & \textit{DS-V3} & \textit{Claude} & \textit{GPT3.5} & \textit{GPT-4} 
    \\
    \midrule
TM & 58.8 & 56.5 & 56.5 & 46.5 & 52.9 & 53.5 & 60.0 & 45.9 & 50.0 & 55.3 \\
DTI & 50.0 & 50.5 & 45.8 & 42.6 & 47.2 & 54.1 & 52.8 & 39.8 & 37.5 & 49.1 \\
PFE & 62.0 & 59.2 & 59.6 & 54.5 & 62.9 & 45.4 & 67.0 & 43.7 & 54.9 & 67.1 \\
PCE & 70.4 & 71.5 & 67.3 & 58.8 & 70.4 & 62.4 & 76.3 & 52.1 & 60.3 & 70.4 \\
FM & 55.9 & 54.6 & 50.5 & 42.6 & 53.8 & 68.9 & 56.2 & 41.5 & 44.8 & 53.0 \\
RM & 20.3 & 18.2 & 20.3 & 19.6 & 16.1 & 21.7 & 23.8 & 25.2 & 9.1 & 21.0 \\
IE & 26.1 & 30.4 & 21.7 & 30.4 & 26.1 & 30.4 & 34.8 & 21.7 & 21.7 & 30.4 \\
\hdashline
\textbf{AVG.} & \textbf{55.7} & 54.8 & 52.6 & 46.5 & \textbf{53.7} & 53.5 & \textbf{58.7} & 43.0 & 45.7 & 55.6 \\
    \bottomrule
    \end{tabular}
    }
    \caption{The table presents the results of various models in the MLDebugging benchmark, including Qwen2.5, Qwen2.5-Coder, Llama3.1, Mistral, closed-source models, and the DeepSeek(DS) R1 distillation series(The models we use are all based on the Instruct version). \textbf{Bolded numbers} indicate the highest pass rate achieved within models of the same parameter size.The Category column represents the initials of the category names listed in Table\ref{tab:count}}
    \label{tab:main-results}
\end{table*}

\subsection{Experimental Settings}
\paragraph{Model Settings}
We conducted experiments on a wide range of models, including both open-source and closed-source systems, to provide a comprehensive understanding of multi-library code debugging.~Specifically, we selected open-source models such as Qwen2.5~\citep{yang2024qwen2}, Qwen2.5-Coder~\citep{hui2024qwen2}, LLama3.1~\citep{LLama3.1}, Mistral~\citep{jiang2024mixtral}, and DeepSeek~\citep{deepseek}, as well as closed-source models including the GPT series~\citep{ChatGPT} and Claude~\citep{anthropic2024claude}. Additionally, we evaluate emerging reasoning models, including the DeepSeek R1 Distill series~\citep{guo2025deepseek} and QwQ-Preview~\citep{QwQ}, to investigate how advanced reasoning capabilities contribute to improved debugging performance.

\paragraph{Metric Settings}
The pass rate represents the proportion of code that pass all test cases. Given test cases \( \mathcal{T}_i \) and the code \( \hat{C}^{i}_{\mathcal{R},l} \) for $i$-th sample, the pass rate can be calculated as:
\begin{equation}
	\mathcal{P} = \frac{1}{n} \sum_{i=1}^{n} (\bigwedge _{t \in \mathcal{T}_i} \left [   \text{\texttt{exec}} (\hat{C}^{i}_{\mathcal{R},l} | t) = \texttt{pass}\right ] ),
\end{equation}
where $n$ denotes the number of codes in benchmark. Here, $ \bigwedge _{t \in \mathcal{T}} [*]$ denotes the logical ``and'' operation across all $t \in \mathcal{T}$. This expression returns 1 if all test cases pass, and 0 otherwise.

\subsection{Main Results}

We evaluate MLDebugging with different LLMs varying in size, including closed- and open-source LLMs ranging from 7B to 72B.

\paragraph{\textit{All LLMs face challenges in MLDebugging.}} To assess the debugging capabilities of LLM in multi-library scenarios, we conduct an in-depth evaluation of various models' performances on MLDebugging.
As shown in Table \ref{tab:main-results}, none of the LLMs demonstrate a significantly high pass rate in ML debugging. Specifically, the highest performance observed on the DeepSeek V3 dataset is only 58.7\%, which indicates that all LLMs face substantial challenges in MLDebugging. This result also highlights significant room for improvement in the models' multi-library debugging capabilities.

\paragraph{\textit{LLMs with increasing parameter sizes exhibit diminishing returns in MLDebugging.}}
To understand the impact of varying model sizes on our benchmark, we evaluate a series of LLMs with different parameter counts.
As shown in Table \ref{tab:main-results}, model performance improves significantly from 7B to 32B parameters. 
However, from 32B to 72B, the performance gain levels off and may even decline. 
This suggests that our benchmark cannot be fully addressed by scaling alone and presents unique challenges inherent to multi-library tasks, 
One key reason is the scarcity of multi-library debugging data in the training corpus \cite{du2025graphmaster}; therefore, scaling the model does not substantially enhance reasoning capabilities, with accuracy remaining around 50–60\% \citep{du2025graphoracle}.
Moreover, direct prompting strategies fall short in overcoming the intricate knowledge dependencies and interactions among libraries during debugging.
These issues highlight inherent difficulties in multi-library tasks that require further research.

\paragraph{\textit{LLMs exhibit varying capabilities across different multi-library debugging categories in MLDebugging.}} To evaluate the debugging abilities of LLMs on different types of bugs, we conduct a detailed experiment based on the classification in Table~\ref{tab:count}.
Table~\ref{tab:main-results} presents the pass rate of various models on MLDebugging. While LLMs are somewhat effective for addressing certain bug types, multi-library debugging continues to pose significant challenges. Specifically, the categories TM \& DTI, which involve \textit{parameter-level} debugging with variable types and specific forms, and the following categories (PFE, PCE, \& FM), which focus on \textit{function-level} debugging, demonstrate relatively better performance. 
In contrast, the last two categories (RM \& IE), which require reasoning and debugging at the \textit{library-level}, show notably lower performance, with a gap of nearly 20\% in pass rates. 
This disparity underscores the varying capabilities of LLMs across different multi-library debugging tasks and highlights the need for targeted improvements, particularly in handling debugging challenges of different complexity.

\subsection{Analysis}
In this section, we thoroughly analyze the behaviors exhibited by LLMs, with a particular focus on examining their performance from the perspective of library-specific understanding.

\begin{figure}[t]
	\includegraphics[width=\columnwidth]{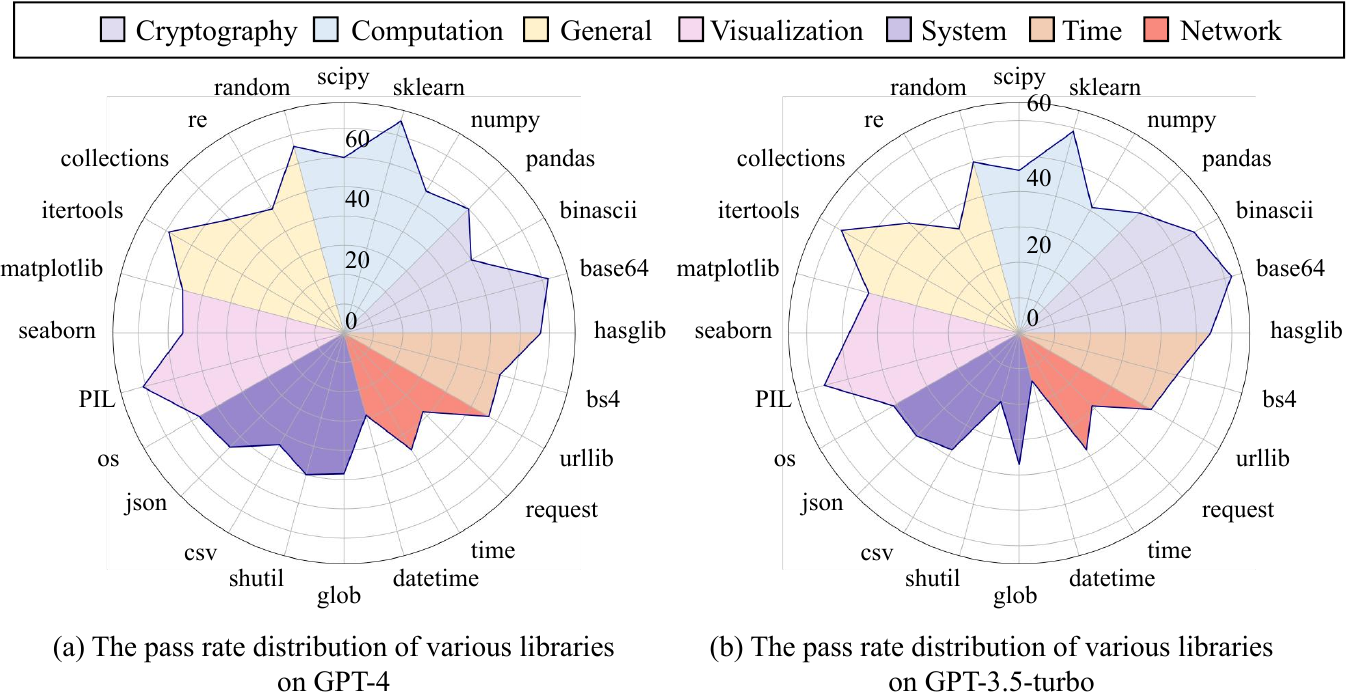}
	\caption{The usage scenarios of Python libraries are categorized into eight distinct domains, several of the most representative libraries are selected.}
	\label{fig:library}
\end{figure}

\subsubsection{Impact of Library Usage Scenario}

To gain a comprehensive understanding of the LLMs' debugging performance across various libraries, we adopt the scenario classification from BigCodeBench~\citep{zhuo2024bigcodebench}. Figure \ref{fig:library} illustrates the debugging pass rates for representative libraries within each scenario. The experimental results indicate that the LLMs' debugging performance varies across different types of libraries, revealing the following insights:

\paragraph{\textit{LLMs perform well in commonly well-regularized and structured libraries.}}
As illustrated in Figure \ref{fig:library}, models perform well in structured and well-regularized libraries covering general algorithms (General), data processing (Computation), and tasks related to encryption and visualization (Visualization). Specifically, they achieve high pass rates in libraries like itertools (0.641), collections (0.570), sklearn (0.654), base64 (0.724), and PIL (0.714) on GPT-4. This strong performance is likely due to the well-structured and clear definition of tasks in large-scale corpora, which allow LLMs to effectively learn and apply general debugging.

\paragraph{\textit{LLMs struggle with dynamic and unstructured multi-library debugging.}} As shown in Figure \ref{fig:library}, LLMs underperform in tasks involving network communication (NetWork) and time processing (Time), particularly when using libraries like bs4 (0.286), urllib (0.375), and nltk (0.167) on GPT-4. These deficiencies in time-related libraries highlight the model’s limitations in dynamic debugging, specifically in its understanding of time logic and cross-timezone processing. This results in challenges when trying to accurately detect anomalous behaviors in dynamic environments.

\subsubsection{Impact of Library Prevalence}

\begin{table}
	\centering
	\begin{tabular}{l c r}
		\toprule
		\textbf{Variable} & \textbf{Correlation} & \textbf{P-Value} \\
		\midrule
		lines of code & -0.0071 & 0.9654 \\
		library count & -0.2113 & 0.1906 \\
        prevalence & 0.4094 & 0.0087 \\
		\bottomrule
	\end{tabular}
	\caption{The table presents the point-biserial correlation coefficients and p-values between these three factors (lines of code, library count, and library prevalence) and the pass/fail outcomes.}
	\label{tab:related}
\end{table}

\paragraph{\textit{High prevalence of libraries elicits the models' capacity on MLDebugging.}} In the experiments presented in Table~\ref{tab:main-results}, LLMs demonstrate poor performance in MLDebugging. To explore the underlying causes of these difficulties, we hypothesized that the challenges in debugging are associated with factors such as code length, the quantity of libraries used, and the prevalence of LLMs encountered on the internet corpus. To test this hypothesis, we computed the correlation between each of these factors and the pass rate. As shown in Table~\ref{tab:related}, we find that the prevalence of libraries exhibits the strongest correlation with the debugging difficulty, suggesting that models tend to perform better when handling libraries that are more commonly encountered.

\subsubsection{Exploration}

\begin{figure}[t]
	\includegraphics[width=\columnwidth]{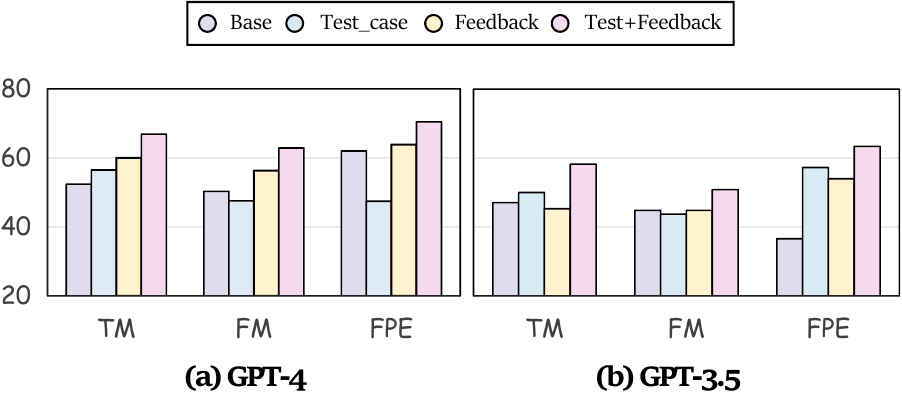}
	\caption{The debugging effect achieved by incorporating runtime information for different bug categories in Table~\ref{tab:count}.}
	\label{fig:exec}
\end{figure}

\paragraph{\textit{Both test cases and runtime error messages are essential for MLDebugging.}}
Execution feedback consistently acts as a vital information source and plays an essential role in effective debugging \citep{zhong-etal-2024-debug}. To explore its impact, we introduce detailed test case and runtime error message information separately into prompts to validate the effectiveness for multi-library debugging. As illustrated in Figure~\ref{fig:exec}, the results illustrate the debugging scenarios based on test cases, runtime feedback, and their combined use. The results demonstrate that providing either test cases or runtime feedback individually offers additional information, thereby improving the model's performance in most cases. However, in certain experiments involving only test case feedback or only runtime feedback, we incorporate both information to the LLMs' inputs, leading to better and more robust performance. This suggests that supplying both comprehensive test cases and runtime error messages is essential for ensuring stable performance improvements in debugging tasks.

\begin{figure}[t]
	\includegraphics[width=\columnwidth]{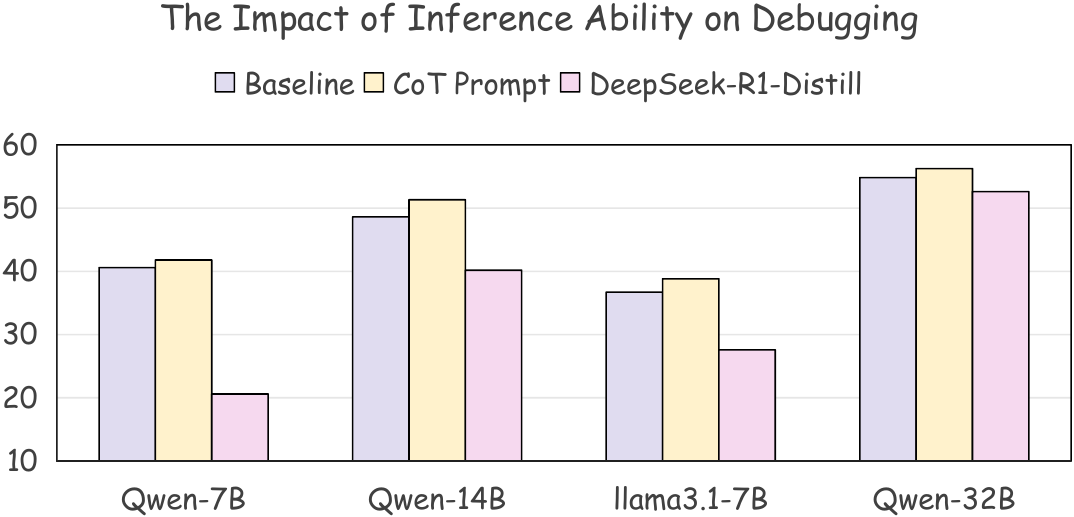}
	\caption{The results of various models under baseline conditions, with CoT prompts, and following R1 distillation training.}
	\label{fig:reasoning}
\end{figure}
\paragraph{\textit{CoT achieves superior performance in MLDebugging.}}
The Chain-of-Thought (CoT) prompt-based approach has been shown to exhibit enhanced reasoning capabilities in various tasks~\citep{wei2022chain, qin2023crosslingualpromptingimprovingzeroshot, zhang2024autocap, chen2024unlocking}. To investigate this effectiveness for MLDebugging, we adopt the approach introduced by \citet{kojima2022large}, wherein the prompt ``Let's think step by step'' is used to trigger LLMs to generate a CoT reasoning process. Specifically, we assess the performance of contemporary LLMs with parameter sizes ranging from 7B to 32B, including Qwen2.5 and LLaMA 3.1. As shown in Figure~\ref{fig:reasoning}, the CoT prompt-based approach consistently demonstrates a significant improvement in reasoning capabilities, particularly in debugging tasks. These enhanced logical reasoning skills prove to be crucial for improving performance in multi-library debugging scenarios.

\paragraph{\textit{Reasoning models based solely on distillation fail to improve task performance.}} 
Emerging research shows that test-time scaling techniques in reasoning models greatly boost the effectiveness of LLMs~\citep{snell2024scaling,jaech2024openai,guo2025deepseek,chen2025towards,chen2025ecm}. Motivated by this insight, we evaluated the Deepseek-R1-Distill-Qwen models. However, as shown in Figure~\ref{fig:reasoning}, although CoT demonstrates improved performance, reasoning models trained with distilled long CoT data exhibit worse performance. We attribute it to the fact that the use of supervised fine-tuning (SFT) with distilled data alone may not sufficiently enhance the model's capabilities. To achieve a more substantial improvement, further exploration of alternative strategies, such as reinforcement learning (RL), is warranted.

\subsection{Error Analysis}
To understand the key challenges in MLDebugging task, we conduct a comprehensive review of the debugged code output by the LLMs. 
We attribute the primary causes to the following two factors:
\paragraph{Absence of library knowledge}
A key challenge of (LLMs) is their limited knowledge~\citep{huang2024survey}. To examine this issue in the context of multi-library debugging, we conduct a detailed manual analysis of LLM-generated outputs, with particular emphasis on their use of specialized knowledge. As demonstrated in Case 1 of \nameref{app:b}, LLM fails to understand the ``virtual memory'' method in the psutil library, leading to a misidentification that hindered its ability to extract the relevant attributes. This highlights a major limitation of LLMs: a superficial or incomplete understanding of specialized software libraries. As a result, effectively localizing errors and reconstructing faulty code with a sufficient and accurate grasp of the relevant programming libraries present significant and persistent challenges in the field of MLDebugging.

\paragraph{Requirements for efficient cross-library debugging}
When dealing with interactions between multiple libraries, the inherent complexities introduce significant challenges. As illustrated in Case 2 in \nameref{app:b}, the model’s inability to comprehend cross-library variables impedes its capacity to detect redundant operations on the dataframe. The variation in variable types across libraries, such as differing classes and structures, prevents the model from performing fine-grained debugging, resulting in inaccuracies when handling task-specific details. As a result, effectively correcting code involving cross-library interactions also presents significant challenges in MLDebugging.

\section{Related Work}

The rise of large language models (LLMs) has also had a considerable impact on Automated Code Debugging (ACD) tasks, as previous datasets have been vulnerable to data leakage risks \citep{just2014defects4j, lin2017quixbugs}. 
To facilitate a smoother transition from traditional datasets to those suited for LLM evaluations, numerous remarkable contributions have surfaced.
DebugBench \citep{tian2024debugbench} stands as the first dataset designed specifically to assess the debugging capabilities of large models. This work collects data from LeetCode, subsequently introducing bugs via model injection. 
In the realm of APR, xCodeEval \citep{khan2024xcodeeval} has proposed three distinct sub-tasks, spanning multiple programming languages, to comprehensively evaluate a model's ability to repair code. 
Meanwhile, MdEval \citep{liu2024mdeval} adopts a multi-language approach, encompassing a benchmark across 18 programming languages.
In addition, the challenges presented by specific real-world scenarios have prompted research into niche areas, with several efforts concentrating on executable code, data processing, and other specialized contexts \citep{yang2024execrepobench, prenner2023runbugrun, galimzyanov2024drawing}. 

These advancements have undeniably propelled the field forward. However, the majority of these datasets are sourced from algorithmic competition platforms, often overlooking the need for Python multi-library scenarios. Therefore, we have constructed a benchmark specifically aimed at evaluating the debugging abilities of models in the context of multiple libraries, providing an in-depth assessment of their proficiency in both static knowledge comprehension and multi-library code interaction.

\section{Conclusion}
This work introduces MLDebugging, a benchmark designed to assess debugging challenges in multi-library code. We conduct a comprehensive analysis on MLDebugging and the experimental results reveal that the current LLMs still struggle in  multi-library scenario.
This work emphasizes the need for further research to improve LLM performance in multi-library settings and provides insights to guide future developments in this field.

\section{Acknowledgement}
This work was supported by the National Natural Science Foundation of China (NSFC) via grant 62306342. This work was sponsored by the Excellent Young Scientists Fund in Hunan Province (2024JJ4070), the Science and Technology Innovation Program of Hunan Province under Grant 2024RC3024 and CCF-Zhipu Large Model Innovation Fund (NO.CCFZhipu202406). We are grateful for resources from the High Performance Computing Center of Central South University.

\section*{Limitations}
We propose MLDebugging, the first multi-library code debugging benchmark, yet there are still two main limitations:
(1) Although the analysis results indicate that the data distribution of MLDebugging closely resembles real-world data, most of the data in MLDebugging are automatically generated by models,  which means there will still be some differences compared to real data. In the future, we consider incorporating more real-world data to further enhance the realism and usability of MLDebugging.
(2) Despite that our experiments comprehensively evaluate various models and error categories, the entire evaluation process can be cumbersome due to the need to configure several external dependencies and complex environments, which consumes a significant amount of time.

\bibliography{custom}
\bibliographystyle{acl}
\clearpage

\appendix

\section*{Appendix}
\section{Dataset Construction Details}
\label{sec:appendixA}

\subsection{Prompts for Dataset Construction}
This appendix includes the prompts used during both the dataset construction phase and the evaluation phase.

In the second stage of dataset construction, the prompts provided to the model for debugging tasks include several key components: the instruction, the buggy code, the test case, the correct code from the original dataset, and the error message. The model is then required to format and output the solution accordingly.we present the prompts used during the second stage of dataset construction, which are provided to the model for debugging tasks.

\begin{tcolorbox}
	\small
	We will provide a Python code snippet, task\_func, which involves multiple libraries and contains bugs. Your task is to debug the code, identify the issues across the libraries, and categorize the bugs. Please follow these steps:
	Review the requirements of the code and thoroughly analyze the task\_func.\\
	Using the provided test cases and error messages, identify and list the bugs in the code, along with a description of each issue.
	Correct each identified bug, provide the updated code, and categorize the bugs according to the relevant multi-library issue types.
	The following information is provided for debugging:\\
	\textcolor{blue}{<instruct>}: The task of this code segment.\\
	\textcolor{blue}{<bug\_code>}: The code that needs debugging.\\
	\textcolor{blue}{<canonical\_solution>}: The corrected version of the code.\\
	\textcolor{blue}{<test\_case>}: Sample test cases for validation.\\
	\textcolor{blue}{<error>}: Relevant error messages.\\
	
	The identified bugs are categorized as follows:
	{}
	
	Please output each identified bug in the following format, Write each part's content with only one tag, Each part must be included:
	
	\textcolor{blue}{<bug\_des>}\\
	Detailed description of the bug
	\textcolor{blue}{</bug\_des>}
	
	\textcolor{blue}{<code>}
	\# Import necessary package(s) and provide the refined function \\code without comments\\
	import ……\\
	def task\_func(\\
	\textcolor{blue}{</code>}
	
	\textcolor{blue}{<category>}
	Only output the category names from the seven categories mentioned above
	\textcolor{blue}{</category>}
	
	Few-Shot:
	{}
\end{tcolorbox}

we include the prompts employed during the data generation process, covering the analysis and bug injection phases.

\begin{tcolorbox}
	\small
	You will receive a piece of code that is a function designed with multiple packages, and its corresponding multi-library AST structure is also provided. Your tasks are as follows: \\
	(1) Analyze the relationships between multiple libraries from the perspective of variable passing, based on the provided code and its corresponding multi-library AST (Abstract Syntax Tree) structure.
	(2) Inject a specific type of bug, provide a description of the bug, and the code where the bug is injected (do not include any comments).
	Type Description:
	{}
	
	The input format is as follows:\\
	\textcolor{blue}{<Instruction>}
	Code Requirements
	\textcolor{blue}{</Instruction>}

	\textcolor{blue}{<Original\_Code>}
	Correct Implementation
	\textcolor{blue}{</Original\_Code>}
	
	\textcolor{blue}{<AST>}
	Abstract Syntax Tree
	\textcolor{blue}{</<AST>}
	
	The output format is as follows:\\
	\textcolor{blue}{<AST\_analysis>}
	Analyze the relationships between multiple libraries, focusing on variable passing, Analyze the relationships between multiple libraries from the perspective of variable passing, based on the provided code and its corresponding multi-library AST (Abstract Syntax Tree) structure. the provided code and its corresponding multi-library AST (Abstract Syntax Tree) structure.\\
	\textcolor{blue}{</AST\_analysis>}
	
	\textcolor{blue}{<bug\_des>}
	This is a description of multi-database bugs for the specific category required.\\
	\textcolor{blue}{</bug\_des>}
	
	\textcolor{blue}{<bug\_code>}
	\# Import the necessary packages and provide the bug-implanted \\code without comments\\
	import ……\\
	def task\_func(\\
	\textcolor{blue}{</bug\_code>}
\end{tcolorbox}

Finally, we outline the various prompts utilized during the testing phase.
Standard evaluation prompt:
\begin{tcolorbox}
	\small
	There is an important info hidden inside a lot of irrelevant text. Find it and memorize them. I will quiz you about the important information there.\\
	Please review the task\_func function for errors. Begin by reading the provided instructions to understand the intended behavior of the function. Ensure the code aligns with the requirements and identify any issues. Correct any errors found and provide the revised code.
	
	Input format:\\
	\textcolor{blue}{<instruct>}: Code requirements and expected functionality\\
	\textcolor{blue}{<bug \_ code>}: The original (bugged) version of the code.\\
	
	Please output only the debugged code under the label \textcolor{blue}{<corrected\_code>}, without any additional text or comments:\\
	Output Format Example\\
	\textcolor{blue}{<corrected\_code>}\\
	import ……\\
	def task\_func(……\\
	\textcolor{blue}{</corrected\_code>}
\end{tcolorbox}
Reasoning model prompt
\begin{tcolorbox}
	\small
Please review the task\_func function for errors. Begin by reading the provided instructions to understand the intended behavior of the function. Ensure the code aligns with the requirements and identify any issues. Correct any errors found and provide the revised code.

Input format
\textcolor{blue}{<instruct>}: Code requirements and expected functionality
\textcolor{blue}{<bug\_code>}: The original (bugged) version of the code.

Please output the final answer at the end, enclosed in markdown format, without any additional text or comments.
Final answer output Format Example:
```python
import ……
def task\_func(……
```
\end{tcolorbox}
CoT prompt
\begin{tcolorbox}
	\small
Please review the task\_func function for errors. Begin by reading the provided instructions to understand the intended behavior of the function. Ensure the code aligns with the requirements and identify any issues. Correct any errors found and provide the revised code.\\

Input format\\
\textcolor{blue}{<instruct>}: Code requirements and expected functionality\\
\textcolor{blue}{<bug\_code>}: The original (bugged) version of the code.\\

Let's think Step by Step to Solve this problem.
Please output the final answer at the end, enclosed in markdown format, without any additional text or comments.\\
Final answer Output Format Example\\
```python\\
import ……\\
def task\_func(……\\
```
\end{tcolorbox}

\subsection{The Details of Manual Annotation}
To ensure the quality of the dataset, we provided training for the data annotate team using a sample of 50 entries and established a structured workflow for dataset annotation:
\begin{itemize}
	\item [(1)] The dataset is divided, with each annotator assigned an equal portion of the data for labeling.
	\item [(2)] Each annotator first reviews 50 labeled samples, subsequently following the established guidelines for further annotation.
	\item [(3)] It is imperative that the data reviewed each day meets the criteria, with the bug code matching both its category and description, and the "Golden" code passing corresponding test cases.
	\item [(4)] Any unresolved issues are addressed through consensus within the annotator group.
\end{itemize}
The guidelines for the annotate process are as follows:
\begin{tcolorbox}
	\small
	\textbf{Instruction}\\
	To construct a Code Debug dataset, we first need to debug the code containing bugs. The debugging process involves a thorough examination of both the model and the corrected code. The dataset includes the following information: the code's instructions, the bug categories, the buggy code, the golden code, and testcase. These elements provide essential support for subsequent debugging and model evaluation.\\
	\textbf{Workflow}\\
	The provided files contain a jsonl file, which can be processed using the check.py script to generate a consolidated output file for easier inspection. The Check folder holds the necessary information for performing code debugging, while the RUN folder offers convenient debugging support for code testing.
	During the debugging process:
	\begin{itemize}
		\item [(1)]  It is essential to verify whether the buggy code aligns with the bug description and to assess it according to the classification standards provided in the accompanying materials. If discrepancies are found, corrections must be made accordingly.
		\item [(2)]  The golden code must be executed, and its correctness verified through the corresponding test cases. If it fails to pass the test cases, manual corrections are required.
		\item [(3)]  The correction process for each data entry should be thoroughly documented to ensure traceability of the modifications.
	\end{itemize}
\end{tcolorbox}

\section{Error Analysis}
\label{app:b}
\textbf{Case 1 }\\
In the first example, the model fails to locate and refactor the code due to a lack of knowledge regarding the virtual\_memory method in the psutil library. As a result, the outputted repair version remains unchanged.
\begin{tcolorbox}
	\textbf{Buggy Code: }
	\begin{lstlisting}
import psutil
import platform
def task_func():
	system_info = {}
	system_info['OS'] = platform.system()
	system_info['Architecture'] = platform.architecture()[0]
	memory = psutil.virtual_memory()
	system_info['Memory Usage'] = f"{memory.percent}%"
	return system_info
	\end{lstlisting}
\end{tcolorbox}

\begin{tcolorbox}
	\textbf{Repaired Code: }
	\begin{lstlisting}
import psutil
import platform
def task_func():
    system_info = {}
	system_info['OS'] = platform.system()
	system_info['Architecture'] = platform.architecture()[0]
	memory = psutil.virtual_memory()
	system_info['Memory Usage'] = f"{memory.percent}%"
	return system_info
	\end{lstlisting}
\end{tcolorbox}

\begin{tcolorbox}
	\textbf{Golden Code: }
	\begin{lstlisting}
import psutil
import platform
def task_func():
    system_info = {}

	system_info['OS'] = platform.system()
	system_info['Architecture'] = platform.architecture()[0]

	total_memory = psutil.virtual_memory().total
	used_memory = psutil.virtual_memory().used
	system_info['Memory Usage'] = f'{used_memory/total_memory*100:.2f}%'

	return system_info
	\end{lstlisting}
\end{tcolorbox}
\noindent\textbf{Case 2 }\\
In the following error sample, the model fails to understand the input information during execution, leading it to overlook the redundant operation of removing missing values and reconstructing the index.
\begin{tcolorbox}
	\textbf{Buggy Code: }
	\begin{lstlisting}
import pandas as pd
import matplotlib.pyplot as plt
def task_func(data_list):
	df = pd.DataFrame(data_list)
	fig, ax = plt.subplots()
	for column in df:
	
	ax.plot(df[column].dropna().reset_index(drop=True), label=column)
	ax.set_title("Student Scores over Tests")
	ax.set_xlabel("Test Number")
	ax.set_ylabel("Score")
	
	return ax
	\end{lstlisting}
\end{tcolorbox}

\begin{tcolorbox}
	\textbf{Repaired Code: }
	\begin{lstlisting}
import pandas as pd
import matplotlib.pyplot as plt
def task_func(data_list):
	df = pd.DataFrame(data_list)
	fig, ax = plt.subplots()
	for column in df:
	ax.plot(df[column].dropna().reset_index(drop=True), label=column)
	ax.set_title("Student Scores over Tests")
	ax.set_xlabel("Test Number")
	ax.set_ylabel("Score")
	ax.legend()
	return ax
	\end{lstlisting}
\end{tcolorbox}

\begin{tcolorbox}
	\textbf{Golden Code: }
	\begin{lstlisting}
import pandas as pd
import matplotlib.pyplot as plt
def task_func(data_list):
	df = pd.DataFrame(data_list)
	fig, ax = plt.subplots()
	for column in df:
	ax.plot(df[column], label=column)
	ax.set_title("Student Scores over Tests")
	ax.set_xlabel("Test Number")
	ax.set_ylabel("Score")
	
	return ax
	\end{lstlisting}
\end{tcolorbox}

\end{document}